\documentclass[lettersize,journal]{IEEEtran}
\usepackage{amsmath,amsfonts}
\usepackage{algorithmic}
\usepackage{algorithm}
\usepackage{array}
\usepackage[caption=false,font=normalsize,labelfont=sf,textfont=sf]{subfig}
\usepackage{textcomp}
\usepackage{stfloats}
\usepackage{url}
\usepackage{verbatim}
\usepackage{graphicx}
\usepackage{cite}
\usepackage{multirow}
\usepackage{booktabs}
\begin{document}

\title{Stable Relay Learning Optimization Approach \\for Fast Power System Production Cost Minimization Simulation}

\author{Zishan Guo,~\IEEEmembership{Student Member,~IEEE}, Qinran Hu, ~\IEEEmembership{Senior Member,~IEEE}, Tao Qian,~\IEEEmembership{Member,~IEEE}, Xin Fang, ~\IEEEmembership{Senior Member,~IEEE}, Renjie Hu, ~\IEEEmembership{Member,~IEEE}, Zaijun Wu, ~\IEEEmembership{Senior Member,~IEEE}

\thanks{Zishan Guo, Qinran Hu, Tao Qian, Renjie Hu, and Zaijun Wu are with the School of Electrical Engineering, Southeast University, Nanjing 210000, China (e-mail: zishanguo@seu.edu.cn; qhu@seu.edu.cn; taylorqian@seu.edu.cn; hurenjie@seu.edu.cn; zjwu@seu.edu.cn). Xin Fang is with the Electrical and Computer Engineering Department of Mississippi State University, Starkville, MS, USA 39762 (xfang@ece.msstate.edu).}
\thanks{Manuscript received xxx; revised xxx.}}



\maketitle

\begin{abstract}
Production cost minimization (PCM) simulation is commonly employed for assessing the operational efficiency, economic viability, and reliability, providing valuable insights for power system planning and operations. However, solving a PCM problem is time-consuming, consisting of numerous binary variables for simulation horizon extending over months and years. This hinders rapid assessment of modern energy systems with diverse planning requirements. Existing methods for accelerating PCM tend to sacrifice accuracy for speed. In this paper, we propose a \emph{stable relay learning optimization} (\emph{s-RLO}) approach within the Branch and Bound (B\&B) algorithm. The proposed approach offers rapid and stable performance, and ensures optimal solutions. The two-stage \emph{s-RLO} involves an \emph{imitation learning} (\emph{IL}) phase for accurate policy initialization and a \emph{reinforcement learning} (\emph{RL}) phase for time-efficient fine-tuning. When implemented on the popular SCIP solver, \emph{s-RLO} returns the optimal solution up to 2$\times$ faster than the default \emph{relpscost} rule and 1.4$\times$ faster than \emph{IL}, or exhibits a smaller gap at the predefined time limit. The proposed approach shows stable performance, reducing fluctuations by approximately 50\% compared with \emph{IL}. The efficacy of the proposed \emph{s-RLO} approach is supported by numerical results.
\end{abstract}


\section{Introduction}
\label{section 1}
\IEEEPARstart{P}{roduction} cost minimization (PCM) simulation is a prevalent tool for evaluating the economic efficiency and operational reliability of power systems. Typically, PCM problem is framed as amixed-integer linear programming (MILP) problem. However, practical PCM problems often span monthly or yearly time frames, introducing numerous binary variables into the MILP formulation. This abundance of binary variables increases the computational complexity and prolongs the solving time \cite{ref1}. Additionally, as power systems evolve in terms of their structure (e.g., increased contributions from renewables, access to adjustable equipment, flexible grid topologies), they present a multitude of planning options. The long computation times required to solve PCM problems undermine the imperative for rapid assessment of diverse planning alternatives.

Many researchers have proposed strategies to accelerate the solution to PCM problems, including reduction-based \cite{ref2, ref reduction1, ref reduction2, ref reduction3,ref reduction4, ref reduction5,ref reduction6}, relaxation-based \cite{ref relxation1,ref relxation2,ref relxation3}, and partition-based \cite{ref partition1,ref partition2,ref partition3} approaches. However, these approaches may return inaccurate solutions. In particular, reduction-based approaches may fail because the predetermined binary variables contradict temporal constraints \cite{ref2} or nonidentical generators are clustered together \cite{ref reduction4, ref reduction5,ref reduction6}, whereas relaxation-based approaches introduce errors through the extended feasible region of Lagrangian relaxation~\cite{ref relxation1,ref relxation2}, linearization\cite{ref relxation3} and other incorrect approximations. Partition-based approaches may violate their temporal constraints in split sections separated by time-domain methods \cite{ref partition1,ref partition2} or geographic-domain methods \cite{ref partition3}. In addition,  machine learning (ML) techniques have been applied to aid in the solving of PCM problems, such as by predicting initial solutions \cite{ref ML1-1}, determining the state of the binary variables \cite{ref ML2-1,ref ML2-2}, identifying redundant variables and constraints \cite{ref ML3-1,ref ML3-2}. However, the performance of ML-based approaches hinges on the quality of the training and testing datasets, often resulting in infeasible solutions.

Regrettably, the pursuit of an acceptable solution time in existing methodologies is frequently realized at the cost of accuracy. In fact, the requirements for accurate solutions remain essential to facilitate cost-benefit analyses and comparisons amid various planning choices, which in turn inform policy, investment, and regulatory decisions in power systems \cite{ref3}. Thus, to obtain accurate solutions while accelerating the solving process, a perspective rooted in algorithmic enhancements is imperative. The branch and bound (B\&B) algorithm is widely adopted for accurate solutions in MILP problems. Efficient variable selection is crucial in B\&B, and traditional rules such as strong branching and pseudo-cost often result in lengthy solving times, particularly when dealing with large MILP problems. The default \emph{relpscost} rule in the popular SCIP solver is widely used, but incurs a nonnegligible time overhead \cite{ref3.5}. The slow variable selection process hinders further speed improvements in B\&B solving.

Recently, ML has been usd to accelerate PCM solving by focusing on the solution process of the B\&B algorithm \cite{ref BB-1,ref BB-2}. Reference \cite{ref BB-3} surveys learning-based methods to tackle the two most important decisions in the B\&B algorithm. The simplex initialization approach and learning-based strategy for several crucial elements in B\&B are further reviewed in \cite{ref BB-4}. Therefore, one promising means of obtaining fast and accurate PCM solutions is to use ML to accelerate the variable selection process in the B\&B algorithm. Our previous work \cite{ref4} provides a fast PCM simulation method with an optimality guarantee based on imitate the behavior of \emph{relpscost}. Although it reduces the solving time of \emph{relpscost} by more than a third, \emph{imitation learning} (\emph{IL}) is affected by changes in the environment, which limits the potential for further improvements to the solution.

Hence, this paper proposes a \emph{stable relay learning optimization} (\emph{s-RLO}) approach that ensures fast and stable solutions with a guarantee of optimality. The approach utilizes \emph{IL} to facilitate the formulation of an initial policy network in the nascent stages and subsequently integrates  \emph{reinforcement learning} (\emph{RL}) to refine the network through ongoing interaction with the environment. The knowledge distilled through this learning mechanism is robust to environmental changes and provides a judicious selection of variables. The insights gleaned from this paper promise to yield optimal, time-efficient solutions for PCM problems.

The contributions of this paper are threefold:

1) The conventional B\&B algorithm is enhanced by an ML-based variable selection strategy, significantly accelerating the PCM solving process on the open-source SCIP solver;

2) A two-stage \emph{s-RLO} approach is designed to acquire and refine the variable selection strategy, combining an \emph{IL} phase for rapid policy network formation and an \emph{RL} phase for subsequent policy network refinement.

3) The proposed approach obtains fast and optimal solutions. Besides, the \emph{s-RLO} framework exhibits consistent and stable performance, even when faced with changing environmental conditions.

The remainder of this paper is organized as follows. Section \ref{section 2} presents the PCM formulation considered in this paper. Section \ref{section 3} proposes the \emph{s-RLO} approach, which uses \emph{IL} and \emph{RL} to learn the policy network sequentially for the selection of candidate variables. Section \ref{section 4} presents the computational results for the proposed approach. The conclusions to this paper are discussed in Section \ref{section 5}.

\section{Problem Formulation}
\label{section 2}
\subsection{PCM formulation}
In this section, we present the foundational mathematical model applied to a PCM problem in power system. Let $\mathcal{B}$, $\mathcal{G}$, $\mathcal{L}$, $\mathcal{WF}$, and $\mathcal{SF}$ be sets of power nodes, generators, power lines, wind farms, and  solar farms, respectively. The set $\mathcal{T}=\{1,2,...,T\}$ encompasses the hours within the planning horizon denoted as $\mathit{T}$. The binary variable $\mu_{g,t}$ is introduced to signify the operational status (on/off) of generator $g\in\mathcal{G}$ at time $t\in\mathcal{T}$.

The objective function \eqref{eq1} aims to minimize the sum of the generation cost and penalty cost. Constraint \eqref{eq2} ensures the power balance for power node $b$ at time $t$. The generation capacity limits are enforced by constraint \eqref{eq3}. The upward and downward ramping capacity of generation are constrained by \eqref{eq4}--\eqref{eq5}. Constraints \eqref{eq6}--\eqref{eq9} impose minimum durations for both the on and off states. Equation \eqref{eq10} is the bound constraint for transmission line capacity. Constraints \eqref{eq11}--\eqref{eq12} describe the upward and downward spinning reserve capacity limits. Constraints \eqref{eq13}--\eqref{eq14} describe limitations related to renewable energies.
\begin{equation}
\label{eq1}
\min \sum_{t \in \mathcal{T}}\left(\sum_{g \in \mathcal{G}} \lambda_g P_{g, t}+\sum_{wf \in \mathcal{WF}} c_{wf} P_{c, wf, t}+\sum_{sf \in \mathcal{SF}} c_{sf} P_{c, sf, t}\right)
\end{equation}

\noindent subject to:
\begin{equation}
\label{eq2}
\begin{split}
\sum_{g \in \mathcal{G}} G_{g, b} P_{g, t}+\sum_{l \in \mathcal{L}} H_{l, b} P_{l, t}+\sum_{wf \in \mathcal{WF}} W_{wf, b} P_{wf, t}\\+\sum_{sf \in \mathcal{SF}} S_{sf, b} P_{sf, t}=D_{b, t}, \forall b \in \mathcal{B}, \forall t \in \mathcal{T}    
\end{split}
\end{equation}
\begin{equation}
\label{eq3}
\mu_{g, t} P_g^{\min } \leq P_{g, t} \leq \mu_{g, t} P_g^{\max }, \forall g \in \mathcal{G}, \forall t \in \mathcal{T}
\end{equation}
\begin{equation}
\label{eq4}
P_{g, t+1}-P_{g, t} \leq\left(\mu_{g, t+1}-\mu_{g, t}\right) P_g^{\min }+P_g^{\mathrm{up}}, \forall g \in \mathcal{G}, \forall t \in \mathcal{T}
\end{equation}
\begin{equation}
\label{eq5}
P_{g, t}-P_{g, t+1} \leq\left(\mu_{g, t}-\mu_{g, t+1}\right) P_g^{\min }+P_g^{\mathrm{down}}, \forall g \in \mathcal{G}, \forall t \in \mathcal{T}
\end{equation}
\begin{equation}
\label{eq6}
\sum_{k=t+1}^{k^{\mathrm {on }}} \mu_{g, k} \geq T_g^{\mathrm {on }} \left(\mu_{g, t+1}-\mu_{g, t}\right), \forall g \in \mathcal{G}, \forall t \in \mathcal{T}
\end{equation}
\begin{equation}
\label{eq7}
\sum_{k=t+1}^{k^{\mathrm{off}}} \mu_{g, k} \leq T_g^{\mathrm{off}} \left(\mu_{g, t+1}-\mu_{g, t}+1\right), \forall g \in \mathcal{G}, \forall t \in \mathcal{T}
\end{equation}
\begin{equation}
\label{eq8}
k^{\mathrm{on}}=\min \left\{t+T_g^{\mathrm{on}}, T\right\}
\end{equation}
\begin{equation}
\label{eq9}
k^{\mathrm {off }}=\min \left\{t+T_g^{\mathrm {off }}, T\right\}
\end{equation}
\begin{equation}
\label{eq10}
P_{l}^{\min } \leq P_{l,t} \leq P_{l}^{\max }, \forall l \in \mathcal{L}, \forall t \in \mathcal{T}
\end{equation}
\begin{equation}
\label{eq11}
\begin{split}
\sum_{g \in \mathcal{G}} \mu_{g, t} P_g^{\max }+\sum_{wf \in \mathcal{WF}} P_{wf, t}+\sum_{sf \in \mathcal{SF}} P_{sf, t} \geq D_t+R_t^{\mathrm{up}}, \forall t \in \mathcal{T} 
\end{split}
\end{equation}
\begin{equation}
\label{eq12}
\begin{split}
\sum_{g \in \mathcal{G}} \mu_{g, t} P_g^{\min }+\sum_{wf \in \mathcal{WF}} P_{wf, t}+\sum_{sf \in \mathcal{SF}} P_{sf, t}\leq  D_t-R_t^{\text {down }}, \forall t \in \mathcal{T} 
\end{split}
\end{equation}
\begin{equation}
\label{eq13}
P_{c, wf, t}+P_{wf, t}= P_{forecast,wf, t}, \forall t \in \mathcal{T}
\end{equation}
\begin{equation}
\label{eq14}
P_{c, sf, t}+P_{sf, t}= P_{forecast,sf, t}, \forall t \in \mathcal{T}
\end{equation}
\noindent where $P_{g, t}$ is the power output for generator $g$ at time $t$; $P_{c, wf, t}$ and $P_{c, sf, t}$ reflect power curtailment at wind farm $wf$ and solar farm $sf$ at time $t$, respectively; $\lambda_g$ is the marginal generating cost of generator $g$; $c_{wf}$ and $c_{sf}$ are marginal penalty factors for the curtailment of wind farm $wf$ and solar farm $sf$, respectively; $G_{g,b}$, $H_{l,b}$, $W_{wf,b}$, and $S_{sf,b}$ are incidence matrices between generator $g$, power line $l$, wind farm $wf$, solar farm $sf$, and power node $b$, respectively; $P_{l,t}$ is the power flow on line $l$ at time $t$; $P_{wf,t}$ and $P_{sf,t}$ are the actual outputs of wind farm $wf$ and solar farm $sf$, respectively; $D_{b,t}$ is the load demand of power node $b$ at time $t$; $P_g^{\min}$ and $P_g^{\max}$ are the minimum and maximum power outputs for generator $g$; $P_g^{\mathrm{up}}$ and $P_g^{\mathrm{down}}$ are the hourly upward and downward ramping powers for generator $g$; $T_g^{\mathrm{on}}$ and $T_g^{\mathrm{off}}$ are the minimum on and off time periods for generator $g$; $P_l^{\min}$ and $P_l^{\max}$ are the minimum and maximum line powers for line $l$; $R_t^{\text{up}}$ and $R_t^{\text{down}}$ are the upward and downward spinning reserve capacities, respectively; $P_{forecast,wf,t}$ and $P_{forecast,sf,t}$ are the predicted outputs from wind farm $wf$ and solar farm $sf$.

The MILP-based PCM problem involving the binary variables within constraints \eqref{eq3}--\eqref{eq7} presents significant challenges because of its size, especially when considering a long time horizon $T$, such as monthly or yearly periods. Consequently, this large-scale PCM problem is inherently intricate, making the rapid attainment of its optimal solution a complex task.

\subsection{Branch and Bound}
\begin{figure*}[!t]
\centering
\includegraphics[width=7in]{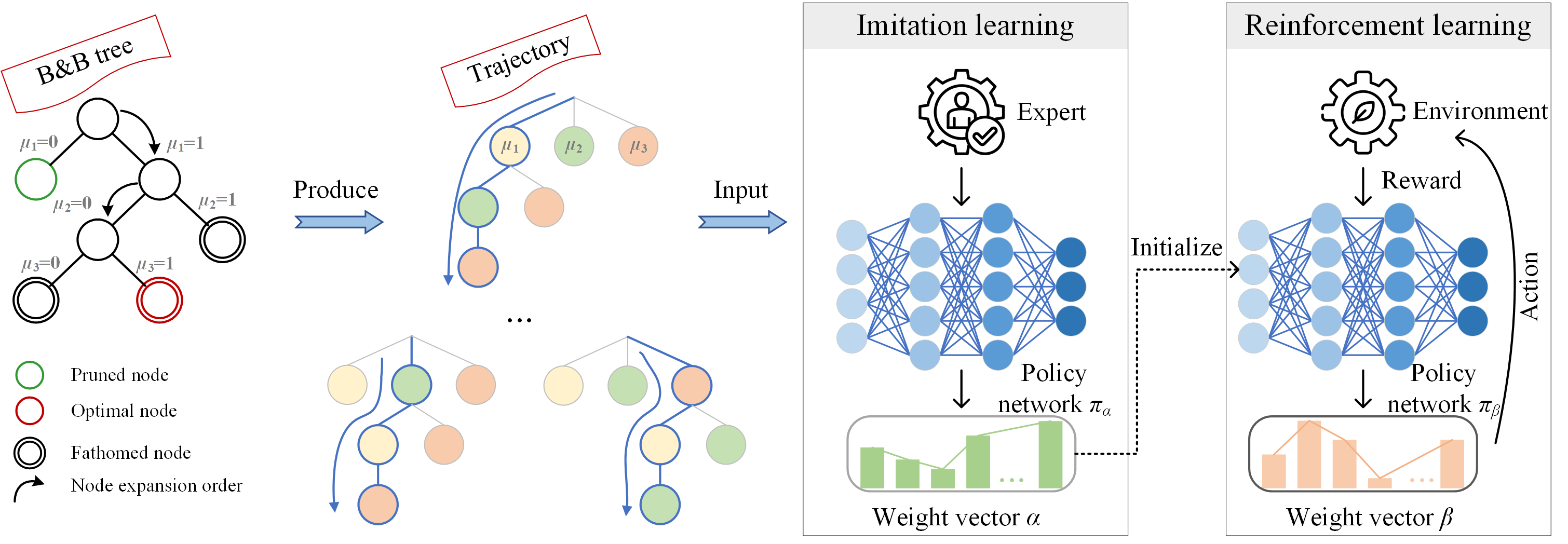}
\caption{Overall framework of proposed \emph{s-RLO} approach.}
\label{fig 2}
\end{figure*}
The B\&B technique operates by iteratively dividing the solution space to identify the optimal solution. The primal bound $z^{\mathrm{primal}}$ and dual bound $z^{\mathrm{dual}}$ are crucial indicators in assessing the progress and completion of the problem-solving process. $z^{\mathrm{primal}}$ represents the best-known feasible solution to the problem, and is initially set to $+\infty$ or is derived from a feasible solution obtained heuristically. As the optimization progresses, $z^{\mathrm{primal}}$ is updated whenever a better feasible solution is discovered. $z^{\mathrm{dual}}$ is obtained by solving a relaxed version of the original problem, and is improved iteratively through enhanced relaxation solutions. The gap, calculated as the difference between $z^{\mathrm{primal}}$ and $z^{\mathrm{dual}}$, reflects the optimality gap. A smaller gap signifies improved solution quality, and a zero gap indicates the discovery of an optimal solution, marking the completion of the problem-solving process.

Node selection and variable selection are two crucial decisions in B\&B. Consider a PCM problem $j$ that achieves its optimal solution after $S_j$ node or variable selection steps. In each step $s\in\mathcal{S}_{j}$ and $\mathcal{S}_{j}=\{1,2,...,S_j\}$, node selection involves choosing a tree node $TN_s$ from the currently open node set, whereas variable selection entails choosing a suitable variable $a_s$ from the set of candidate variables $\mathcal{CV}_s$ associated with node $TN_s$. $\mathcal{CV}_s$ encompasses binary variables with non-integer values in the solution vector $x_s^*$ obtained from linear programming relaxation on $TN_s$. The variables in $\mathcal{CV}_s$ remain unbranched along the path from the root node to $TN_s$. Subsequently, two child nodes are created by branching on $a_s$. Node pruning and bound updating are then conducted based on certain criteria.

The left part of Fig. \ref{fig 2} presents a simplified representation of a B\&B tree for a PCM problem. At every node in the B\&B tree, two distinct bounds are established: the local dual bound, determined through linear programming relaxation, and the local primal bound, extracted from a feasible integer solution within that node. For minimization problems, updates are applied to the global primal and dual bounds whenever local primal and dual bounds are found to be lower than the prevailing global values. To underscore the role of variable selection, the node displaying the greatest potential, as gauged by a lower local dual bound, is selected by default during the node selection phase. Nodes are pruned under the following conditions: 1) if the local dual bound exceeds the current global primal bound and 2) if the subproblem is infeasible. A node is said to be fathomed if it is found to contain a feasible integer solution.

Under these principles, diverse variable selection orders generate B\&B trees with distinct structures. For PCM problems, a judicious choice of the variable selection order is vital in developing a well-structured B\&B tree, compact in both size and height. Such a tree structure will effectively accelerate the identification of optimal solutions compared with alternative selection orders.

Through this analysis, it is evident that variable selection plays a pivotal role in the efficacy of the B\&B approach. Consequently, this paper aims to enhance the efficiency of variable selection. The process involves evaluating candidate variables using a scoring mechanism dictated by the chosen selection rule. Branching is performed on the candidate variable with the highest score, with the aim of curtailing the processing load on nodes. Strong branching is effective in yielding a concise search tree, but has limited practical use because of its substantial computational demands. In practical scenarios, the \emph{relpscost} rule takes precedence, serving as the default strategy in the SCIP solver. Despite its practical utility, \emph{relpscost} can be time-intensive, particularly when confronted with uninitialized and unreliable candidate variables.  Despite this challenge, \emph{relpscost} is the preferred choice as an expert policy, expediting the establishment of an initial policy network. The detailed process is discussed thoroughly in Section \ref{section 3}.


\section{Stable Relay Learning Optimization Approach}
\label{section 3}

This section introduces the  \emph{s-RLO}  approach, which is designed to enhance the selection of variables within the B\&B framework and accelerate the solution of PCM problems in a stable manner. The core concept of this approach is illustrated in Figure \ref{fig 2}.  In this framework,  \emph{IL} leverages insights from the \emph{relpscost} variable selection rule to train a policy network denoted as $\pi_{\alpha}$. The knowledge assimilated through this training enables the policy network to mimic the decision-making process of  \emph{relpscost}. However, \emph{relpscost} might not yield the optimal B\&B tree structure. In addition, the implementation of \emph{IL} does not guarantee perfect imitation accuracy. These limitations impact the effectiveness of \emph{IL}. Thus, the proposed approach integrates \emph{RL}  into the process, continuing the training of the policy network from $\pi_{\alpha}$ to $\pi_{\beta}$. During this phase, a reward mechanism is used to minimize the time required to solve PCM problems. This helps the policy network prioritize time savings over absolute prediction accuracy.

The \emph{s-RLO} approach leverages acquired knowledge instead of relying on \emph{relpscost} to identify a suitable variable for branching. This selection process involves only linear operations, so \emph{s-RLO} has a more straightforward and significantly faster calculation procedure than \emph{relpscost}. Consequently, the \emph{s-RLO} approach demonstrates a promising capacity to accelerate the problem-solving process.

\subsection{Imitation Learning for Policy Network}

The B\&B search can be formulated as a sequential decision-making process that is Markovian in nature \cite{ref5,ref6}. A tuple $\left(F, A, TP, r\right)$ is defined to describe the finite-horizon Markov decision process of B\&B variable selection for PCM problems. $F$ is the state space of the candidate variables. $A$ is the action space for candidate variable selection. $TP$ is the transition probability density function, which generates the probability $TP\left( f_{s+1} \mid f_s, a_s\right)$ that the next state is $f_{s+1}$ given the current state $f_s$ and action $a_s$. $r$ is a reward function. \emph{IL} capitalizes on demonstrations $\mathcal{M}$ acquired through the expertise of \emph{relpscost}, rather than relying on the reward $r$. When tackling a PCM problem $j$ with the expert \emph{relpscost}, a trajectory $\tau_j$ unfolds. Collectively, $\mathcal{M}$ forms a repository of trajectories $\mathcal{M}=\left\{\tau_1,\tau_2,\ldots, \tau_j,\ldots\right\}$, each embedding a sequence of state-action pairs $\tau_j=\left\{f_{j,0}, a_{j,0},\ldots,f_{j,s},a_{j,s}\ldots\right\}$. These state-action pairs serve as inputs to the policy network and compose the variable feature matrix $f_{j,s}=\left\{V_{j,s}, L_{j,s}\right\}$. Here, $V_{j,s}$ includes the numerical attributes of candidate variables on node $TN_s$, such as primal/dual bounds, solution values, and performance statistics from previous branchings. $L_{j,s}$ records the position attributes of node $TN_s$, including the depth, evolution of bounds, and statistics on feasible solutions. The decision label $a_{j,s}$ logs the strategic choice of candidate variable on node $TN_s$. 
 
The policy function $\pi\left(a \mid f\right)$ is a probability density function. It takes the state $f$ as input and outputs the probabilities for all actions. A policy network $\pi_\alpha\left(a \mid f;\alpha\right)$ is used in \emph{IL} to approximate $\pi\left(a \mid f\right)$. The goal of \emph{IL} is to learn a policy network $\pi_\alpha$ with a trainable weight vector $\alpha$ that mimics the state-action pairs in $\mathcal{M}$. The architecture of the policy network $\pi_\alpha$ is referred to \cite{ref7}. Interconnected layers with weight vector $\alpha$ and rectifier nonlinearities constitute the policy network $\pi_\alpha$. The training process of the weight vector $\alpha$ involves maximizing the likelihood of action $a_{j,s}$ given state $f_{j,s}$. Prior to \emph{IL} training, the collected state-action pairs are organized into batches, with each batch containing $BS$ state-action pairs. During each iteration $i$, the batch loss $\mathcal{L}oss_{BS,i}$ is computed using the cross-entropy loss, as specified in \eqref{IL_1}. The weight vector $\alpha$ is then updated based on the average cross-entropy loss in an epoch, employing stochastic gradient descent as shown in \eqref{IL_2}. The final step involves a softmax layer that generates a probability distribution across all candidate variables in $\mathcal{CV}_s$. We summarize this training procedure in Algorithm \ref{alg:IL}.
\begin{algorithm}[t]
\caption{Learning from \emph{relpscost} by \emph{IL}.}\label{alg:IL}
\renewcommand{\algorithmicrequire}{\textbf{Input:}}
\renewcommand{\algorithmicensure}{\textbf{Output:}}
\begin{algorithmic}[1]
\REQUIRE 
A set of state-action pairs; epoch $E$, batch size $BS$; iteration $I$.
\ENSURE 
Trained \emph{IL} policy network $\pi_\alpha\left(a \mid f;\alpha\right)$.
\STATE  Initialize weight vector $\alpha$ randomly.
\FOR{epoch $e =1,2,...,E$}
\FOR{iteration $ i =1,2,...,I$}
\FOR{$s \in\mathcal{BS}_{i}$}
\STATE Use current $\pi_\alpha\left(a \mid f;\alpha\right)$ to yields a probability distribution over all candidate variables $\pi_\alpha\left(\cdot \mid f_{i,s};\alpha\right)$ given state $f_{i,s}$.
\STATE Calculate batch loss $\mathcal{L}oss_{BS,i}$ according to the probability of $a_{i,s}$ being selected given state $f_{i,s}$ as shown in \eqref{IL_1}.
\ENDFOR
\STATE Update weight vector $\alpha$ according to the average cross-entropy loss in epoch $e$ using stochastic gradient descent as shown in \eqref{IL_2}.
\ENDFOR
\ENDFOR
\end{algorithmic}
\end{algorithm}

\begin{equation}
\label{IL_1}
\mathcal{L}oss_{BS,i}=\frac{1}{BS} \sum_{s \in\mathcal{BS}} \log \pi_\alpha\left(a_{i,s} \mid f_{i,s}, \alpha\right)
\end{equation}
\begin{equation}
\label{IL_2}
\alpha \leftarrow \alpha- \frac{1}{\sum_{i \in \mathcal{I}} |\mathcal{BS}_{i}|} \sum_{i \in\mathcal{I}} \sum_{s \in\mathcal{BS}_{i}} \frac{\partial\log \pi_\alpha\left(a_{i,s} \mid f_{i,s}, \alpha\right)}{\partial \alpha}
\end{equation}
\noindent where $\mathcal{BS}=\{1,2,...,BS\}$ represents the set of steps within $BS$; $ \pi_\alpha\left(a_{i,s} \mid f_{i,s}, \alpha\right)$ denotes the probability of selecting $a_{i,s}$ given state $f_{i,s}$ using weight vector $\alpha$ for iteration $i$.

The policy network $\pi_{\alpha}$ acquired through \emph{IL}  is adept at making decisions on candidate variable selection by emulating the behavior of \emph{relpscost}. In practical scenarios, gathering complete sets of states and corresponding actions is challenging. Training data typically only include a fraction of the complete states and corresponding actions. If the current state is present in the training data, $\pi_{\alpha}$ can effectively imitate \emph{relpscost} for a sound choice. However, in cases where the current state is absent from the training data, the decision made by $\pi_{\alpha}$ may be erroneous. Additionally, the cumulative impact of errors across different states could potentially lead $\pi_{\alpha}$ to unexplored trajectories. This situation arises when the policy network makes decisions for states that \emph{relpscost} has never experienced, and $\pi_{\alpha}$ has never been trained on. In these unexplored states, the behavior of the policy network becomes undefined, potentially culminating in prolonged problem-solving times compared with \emph{relpscost}.


\subsection{Reinforcement Learning for Policy Network}

Sharing an identical structure with the policy network $\pi_{\alpha}$, an \emph{RL} enhanced policy network $\pi_{\beta}\left(a \mid f;\beta\right)$ is evolved from $\pi_{\alpha}\left(a \mid f;\alpha\right)$. With the initial weight vector $\alpha$ from $\pi_{\alpha}$, policy network $\pi_{\beta}$ establishes its weight vector $\beta$.

We express the discounted return $u_s$ as shown in \eqref{RL_1}. The actor-value function $Q_{\pi_\beta}\left(f_s, a_s\right)$ is the conditional expectation of $u_s$, as shown in \eqref{RL_2}. The state-value function $V_{\pi_\beta}\left(f_s\right)$ is the expectation of $Q_{\pi_\beta}$. If $A \sim \pi_\beta\left(\cdot \mid f_s\right)$ is discrete, $V_{\pi_\beta}\left(f_s\right)$ can be expressed as \eqref{RL_3}. Policy network $\pi_\beta\left(a \mid f;\beta\right)$ is used to further approximate $\pi\left(a \mid f\right)$ from $\pi_\alpha\left(a \mid f;\alpha\right)$. Similarly, $V_{\pi_\beta}\left(f_s\right)$ can be approximated by $V_{\pi_\beta}\left(f_s;\beta\right)$, as shown in \eqref{RL_4}.

\begin{equation}
\label{RL_1}
u_s=r_s+\gamma r_{s+1}+\cdots+\gamma^{(S_j-s)}r_{S_j}
\end{equation}
\begin{equation}
\label{RL_2}
Q_{\pi_\beta}\left(f_s, a_s\right)=\mathbb{E}\left[u_s \mid f_s, a_s\right]
\end{equation}
\begin{equation}
\begin{aligned}    V_{\pi_\beta}\left(f_s\right)&=\mathbb{E}_A\left[Q_\pi\left(f_s, A\right)\right]\\
&=\sum_{a_s} \pi\left(a_s \mid s_t\right) \cdot Q_{\pi_\beta}\left(f_s, a_s\right)
\end{aligned}
\label{RL_3}
\end{equation}
\begin{equation}
\label{RL_4}
V_{\pi_\beta}\left(f_s ; \beta\right)=\sum_{a_s} \pi\left(a_s \mid f_s ; \beta\right) \cdot Q_{\pi_\beta}\left(f_s, a_s\right)
\end{equation}

$V_{\pi_\beta}\left(f_s;\beta\right)$ can be used to evaluate the quality of state $f_s$ and policy network $\pi_\beta\left(a \mid f;\beta\right)$. Given $f_s$, a better policy network $\pi_\beta\left(a \mid f;\beta\right)$ will give a larger value of $V_{\pi_\beta}\left(f_s;\beta\right)$. To learn $\beta$, we apply a policy-based strategy that maximizes $Z(\beta)$, which is the expectation of $V_{\pi_\beta}\left(f_s;\beta\right)$. The expression of $Z(\beta)$ is shown in \eqref{RL_5}. A policy gradient is used to improve $\beta$ according to \eqref{RL_6}, where $\eta$ is the learning rate. The derivative of $V_{\pi_\beta}(f_s ; \beta)$ is given in \eqref{RL_7}.
\begin{equation}
\label{RL_5}
Z(\beta)=\mathbb{E}_{F}[V_{\pi_\beta}(F ; \beta)]
\end{equation}
\begin{equation}
\label{RL_6}
\beta \leftarrow \beta+\eta \frac{\partial V_{\pi_\beta}(f_s ; \beta)}{\partial \beta}
\end{equation}
\begin{equation}
\begin{aligned}
\frac{\partial V_{\pi_\beta}(f_s ; \beta)}{\partial \beta} & =\sum_{a_s} \frac{\partial \pi(a_s \mid f_s ; \beta)}{\partial \beta}  Q_{\pi_\beta}(f_s, a_s) \\
& =\sum_{a_s} \pi(a_s \mid f_s ; \beta) \frac{\partial \log \pi(a_s \mid f_s ; \beta)}{\partial \beta}  Q_{\pi_\beta}(f_s, a_s) \\
& =\mathbb{E}_A\left[\frac{\partial \log \pi(A \mid f_s ; \beta)}{\partial \beta}  Q_{\pi_\beta}(f_s, A)\right]
\end{aligned}
\label{RL_7}
\end{equation}

However, it is difficult to obtain an accurate assessment of the quality of each variable selection step during the process of solving a PCM problem. That is, we cannot determine the exact value of $Q_{\pi_\beta}(f_s, a_s)$ at every step $s$. In practice, once a PCM problem has been solved, the variable selection steps during this process can be roughly evaluated based on the overall solving time. Therefore, an \emph{RL} algorithm is used to approximate $Q_{\pi_\beta}(f_s, a_s)$.  The relationship between $Q_{\pi_\beta}(f_s, a_s)$ and $u_s$ in \eqref{RL_2} allows us to use observation $u_s$ to approximate $Q_{\pi_\beta}(f_s, a_s)$. Thus, \eqref{RL_7} can be rewritten as follows:
\begin{equation}
\frac{\partial V_{\pi_\beta}(f_s ; \beta)}{\partial \beta} =\mathbb{E}_A\left[\frac{\partial \log \pi(A \mid f_s ; \beta)}{\partial \beta}  u_A\right]
\label{RL_8}
\end{equation}

The expectation of $\frac{\partial \log \pi(a_s \mid f_s ; \beta)}{\partial \beta}  u_s$ cannot be easily calculated. Thus, we randomly sample an action $\hat{a}$ according to the probability density function $\pi(\cdot \mid f_s ; \beta)$. We define $h(\hat{a},\beta)$ as in \eqref{RL_9}. Obviously, the expectation of $h(\hat{a},\beta)$ is equal to $\frac{\partial V_{\pi_\beta}(f_s ; \beta)}{\partial \beta} $, which is written as \eqref{RL_10}. $h(\hat{a},\beta)$ is an unbiased estimate of $\frac{\partial V_{\pi_\beta}(f_s ; \beta)}{\partial \beta} $. Thus, \eqref{RL_6} can be rewritten as \eqref{RL_11}.

\begin{equation}
\label{RL_9}
h(\hat{a},\beta)=\frac{\partial \log \pi(\hat{a} \mid f_s ; \beta)}{\partial \beta}  u_{\hat{a}}
\end{equation}
\begin{equation}
\label{RL_10}
\mathbb{E}_A\left[h(A,\beta)\right]=\frac{\partial V_{\pi_\beta}(f_s ; \beta)}{\partial \beta}
\end{equation}
\begin{equation}
\label{RL_11}
\beta \leftarrow \beta+\eta \frac{\partial \log \pi(\hat{a} \mid f_s ; \beta)}{\partial \beta}  u_{\hat{a}}
\end{equation}

\begin{algorithm}[t]
\caption{\emph{RL} training procedure.}\label{alg:RL}
\renewcommand{\algorithmicrequire}{\textbf{Input:}}
\renewcommand{\algorithmicensure}{\textbf{Output:}}
\begin{algorithmic}[1]
\REQUIRE 
Various PCM problems; epoch $E$; iteration $I$; coefficient for reward $\lambda$ .
\ENSURE 
Trained \emph{RL} policy network $\pi_\beta\left(a \mid f;\beta\right)$.
\STATE  Initialize weight vector $\beta$ with trained $\alpha$.
\FOR{epoch $ e =1,2,...,E$}
\FOR{iteration $ i =1,2,...,I$}
\FOR{problem $j \in\mathcal{Q}_{MB,i}$}
\STATE Use policy network $\pi_\beta\left(a \mid f;\beta\right)$ with latest weight vector $\beta$ to solve problem $j$ for its trajectory $\tau_j$.
\FOR{$s \in\mathcal{S}_{j}$}
\STATE Calculate the return of every step $u_{j,s}$ according to \eqref{RL_12}--\eqref{RL_13}.
\ENDFOR
\ENDFOR
\STATE Update weight vector $\beta$ using gradient ascent, as described in \eqref{RL_14}.
\ENDFOR
\ENDFOR
\end{algorithmic}
\end{algorithm}

The reward $r_{j,s}$ for all non-terminal candidate variable selection steps in trajectory $\tau_j$ of PCM problem $j$ is set to zero, while the reward for the terminal step is determined based on the percentage improvement in the \emph{RL} solving time relative to \emph{relpscost}, as defined in \eqref{RL_12}. If the solving time is unsatisfactory, every step $s$ in the solving process is uniformly treated as improved; conversely, if the solving time is satisfactory, each step $s$ is uniformly considered as unimproved. Setting $\eta = 1$ and using the definition of the return in \eqref{RL_1}, the expression for the return at any step $s\in\mathcal{S}_{j}$ is given by \eqref{RL_13}.
\begin{equation}
\label{RL_12}
r_{j,s}=\left\{\begin{array}{l}
0, s \in\{1, \ldots, S_j-1\} \\
\lambda \frac{ ST_{\text {relpscost },j}-ST_{\mathrm{RL},j}}{ST_{\text {relpscost },j}}, s=S_j
\end{array}\right.
\end{equation}
\begin{equation}
\label{RL_13}
u_{j,s}=\lambda \frac{ST_{\text {relpscost },j}-ST_{\mathrm{RL},j}}{ST_{\text {relpscost },j}}
\end{equation}

\noindent where $\lambda$ represents the coefficient for the reward on the terminal step; $ST_{\mathrm{relpscost},j}$ and $ST_{\mathrm{RL},j}$ denote the solving times for PCM problem $j$ using \emph{relpscost} and \emph{RL}, respectively.

There will be considerable variations in the solving times among randomly generated PCM problems. Thus, relying on the solving trajectory of a single problem to update the weight vector $\beta$ is unreliable. Instead, the weight vector is updated using the trajectories of a mini-batch of problems $\mathcal{Q}_{MB}$ in a direction that maximizes the expected outcome, as illustrated in \eqref{RL_14}. The \emph{RL} training procedure is summarized in Algorithm \ref{alg:RL}.

\begin{equation}
\label{RL_14}
\beta \leftarrow \beta+ \frac{1}{\sum_{j \in \mathcal{Q}_{MB}} |\mathcal{S}_{j}|}\sum_{j \in \mathcal{Q}_{MB}} \sum_{s \in\mathcal{S}_{j}} \frac{\partial \log \pi\left(a_{j,s} \mid f_{j,s}, \beta\right)}{\partial \beta} u_{j,s}
\end{equation}

\begin{figure}[!t]
\centering
\includegraphics[width=2.8in]{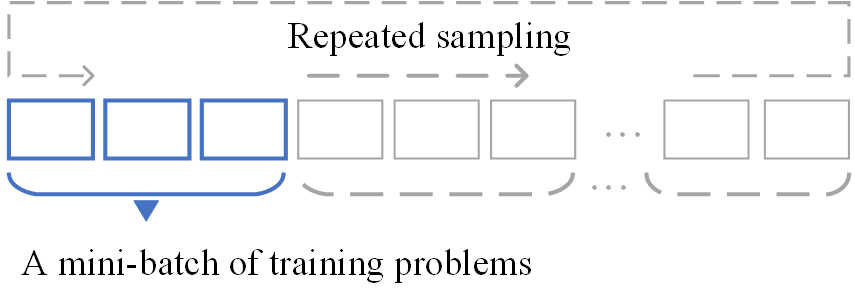}
\caption{Procedure for recycling training problems.}
\label{fig 3}
\end{figure} 

The \emph{RL} network processes PCM problems sequentially. The updated weight vectors used for subsequent problems might not seamlessly apply to preceding ones. As \emph{RL} requires a substantial volume of training problems, generating and solving a large number of these problems using SCIP incurs a significant time cost. To manage this, the training problems are repetitively solved in accordance with the latest weight vector of the \emph{RL} policy network. This procedure is illustrated in Fig. \ref{fig 3}. Repetitive solving involves applying the policy network with the most recent weight vector $\beta$ to solve PCM problems for trajectories. The rewards acquired from these trajectories guide the iterative update of the weight vector $\beta$. This iterative process continues until a proficient policy network $\pi_{\beta}$ has been trained.

Given that rewards are instrumental in shaping the updates to the weight vector $\beta$, policy network  $pn_{\beta}$ typically exhibits enhanced performance compared with $pn_{\alpha}$ in instances involving states that neither $pn_{\alpha}$ nor \emph{relpscost} have encountered during training. However, \emph{IL} occasionally outperforms $pn_{\beta}$ in situations involving states that have previously been encountered. Hence, both $pn_{\beta}$ and $pn_{\alpha}$ can be applied in parallel to address PCM problems. Once the policy network identifies the optimal solution, the problem-solving process is immediately halted.

\section{Case Study}
\label{section 4}
\subsection{Parameter Setting}

\begin{table}[!t]
  \centering
  \caption{Parameters for the generators}
  \setlength{\tabcolsep}{1mm}{
    \begin{tabular}{ccccccc}
    \toprule
    Generators &$P_g^{\max }$(MW) &$P_g^{\min }$(MW) & $P_g^{\mathrm{up}}$(\%) & $P_g^{\mathrm{down}}$(\%) & $P_g^{\mathrm{on}}$(h) & $P_g^{\mathrm{off}}$(h) \\
    \midrule
    G1    & 380   & 76    & 50    & 50    & 2     & 2 \\
    G2    & 775   & 155   & 40    & 40    & 2     & 3 \\
    G3    & 600   & 120   & 35    & 35    & 3     & 3 \\
    G4    & 500   & 100   & 50    & 50    & 2     & 2 \\
    G5    & 900   & 180   & 32    & 32    & 3     & 3 \\
    \bottomrule
    \end{tabular}}%
  \label{tab:table1}%
\end{table}%

\begin{table}[!t]
\caption{Size of PJM 5-bus system problems\label{tab:table2}}
\centering
\begin{tabular}{cccc}
\toprule
$T$ (h) & \# of Cont. var. & \# of Bin. var. & \# of Const.\\
\midrule
336 & 5040 & 1680 & 18761 \\
720 & 10800 & 3600 & 40265 \\
1440 & 21600 & 7200 & 80585 \\
\bottomrule
\end{tabular}
\end{table}

In this section, we present computational results achieved using the proposed \emph{s-RLO} approach. Our implementation of \emph{s-RLO} uses the open-source solver SCIP 8.0.1, and the implementation is carried out in Python 3.6. All computations have been executed on a workstation equipped with an Intel Xeon CPU and an NVIDIA RTX2080Ti GPU.

\subsection{Results Analysis on PJM 5-Bus System }

The effectiveness of the proposed method is evaluated using a PJM 5-bus system, which comprises five generators, a wind farm, and a solar farm. This system has an aggregate generation capacity of 3155 MW. The parameters for the five generators are detailed in Table \ref{tab:table1}. To assess the performance, we formulate three distinct cases in the PJM 5-bus system, and separately consider time horizons of 336 h, 720 h, and 1440 h; the time interval is set to 1 h 

The load and renewable profiles are derived from actual data in a Chinese province, scaled down to 10\% for the experiments. A total of 4500 PCM problems are generated for each case by randomly multiplying the original profiles with sets of normal random noise. The sizes of the PCM problems for 336-, 720-, and 1440-h time horizons are presented in Table \ref{tab:table2}. These generated PCM problems are split into 2000 training problems for \emph{IL}, 2000 training problems for \emph{RL}, and 500 test problems. During the data collection process,  SCIP is employed to solve these problems using the default \emph{relpscost} rule. The solving time limit is set to 0.5 h for each problem.

\begin{figure}[!t]
\centering
\includegraphics[width=3.3in]{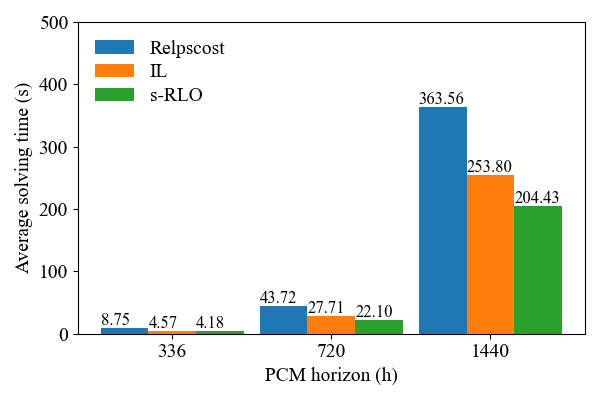}
\caption{Average solving times in PJM 5-bus system.}
\label{fig 5}
\end{figure}

\begin{figure}[!t]
\centering
\includegraphics[width=3in]{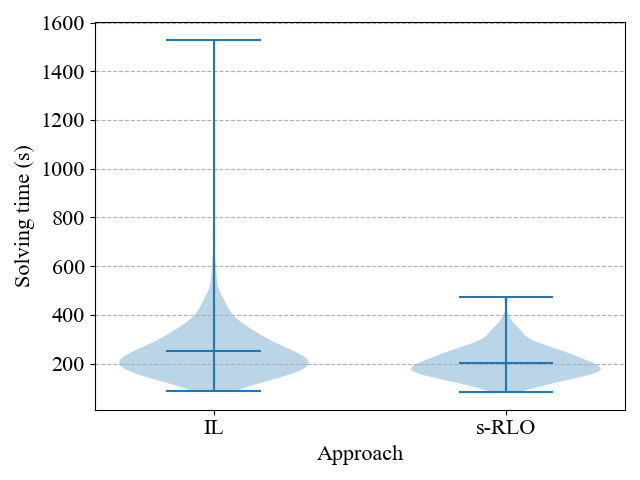}
\caption{Distribution of solving times for 1440-h problems}
\label{fig 6}
\end{figure}

Figure \ref{fig 5} displays the average solving times across all test problems with varying time scales in the PJM 5-bus system. Notably, the use of \emph{IL} produces a significant reduction of more than 30\% in the average solving time compared with the default \emph{relpscost}. Specifically, for problems with a 336-h horizon, \emph{IL} excels, achieving an impressive reduction of approximately 50\% in average solving time relative to \emph{relpscost}. The proposed \emph{s-RLO} approach further shortens the solving time on the basis of \emph{IL}. For example, for the 720- and 1440-h horizon problems, \emph{s-RLO} reduces the solving time by more than 20\% compared with \emph{IL}.

Additionally, \emph{s-RLO} demonstrates more stable performance than \emph{IL} across different test problems. For the 1440-h horizon problems, Figure \ref{fig 6} displays the distribution of solving times, highlighting the considerable variability observed with \emph{IL}. The solving time varies greatly when solving problems with the same horizon. Whereas, the variance of the solving time by applying \emph{s-RLO} is reduced by more than 50\% on average compared with \emph{IL} in different time horizons. In summary, \emph{s-RLO} produces a more stable and superior acceleration effect than \emph{IL} on problems of different scales, achieving acceleration of up to 2$\times$ and 1.25$\times$ compared with \emph{relpscost} and \emph{IL}, respectively, and reducing the variance of the solving times by a factor of 2 compared with \emph{IL}.

\begin{table}[!t]
  \centering
  \caption{Evaluation on randomly selected test problems}
    \begin{tabular}{ccccc}
    \toprule
    Problems & Approach & Solving time (s) & Status & $N_{EP}$ \\
    \midrule
    \multirow{3}[2]{*}{TP-1a} & \textit{Relpscost} & 351.20 & Optimal  & 592 \\
          & \textit{IL} & 250.95 & Optimal  & 350 \\
          & \textit{s-RLO} & 219.55 & Optimal  & 662 \\
    \midrule
    \multirow{3}[2]{*}{TP-2a} & \textit{Relpscost} & 337.48 & Optimal  & 2775 \\
          & \textit{IL} & 278.31 & Optimal  & 1243 \\
          & \textit{s-RLO} & 157.84 & Optimal  & 55 \\
    \midrule
    \multirow{3}[2]{*}{TP-3a} & \textit{Relpscost} & 443.84 & Optimal  & 3271 \\
          & \textit{IL} & 924.73 & Optimal  & 8444 \\
          & \textit{s-RLO} & 248.86 & Optimal  & 1387 \\
    \midrule
    \multirow{3}[2]{*}{TP-4a} & \textit{Relpscost} & 447.22 & Optimal  & 34 \\
          & \textit{IL} & 152.02 & Optimal  & 33 \\
          & \textit{s-RLO} & 152.02 & Optimal  & 33 \\
    \midrule
    \multirow{3}[2]{*}{TP-5a} & \textit{Relpscost} & 291.06 & Optimal  & 1153 \\
          & \textit{IL} & 300.80 & Optimal  & 2448 \\
          & \textit{s-RLO} & 169.06 & Optimal  & 656 \\
    \midrule
    \multirow{3}[2]{*}{TP-6a} & \textit{Relpscost} & 239.98 & Optimal  & 12 \\
          & \textit{IL} & 334.09 & Optimal  & 64 \\
          & \textit{s-RLO} & 285.48 & Optimal  & 282 \\
    \midrule
    \multirow{3}[2]{*}{TP-7a} & \textit{Relpscost} & 415.17 & Optimal  & 3093 \\
          & \textit{IL} & 305.16 & Optimal  & 870 \\
          & \textit{s-RLO} & 293.97 & Optimal  & 187 \\
    \midrule
    \multirow{3}[2]{*}{TP-8a} & \textit{Relpscost} & 537.84 & Optimal  & 3626 \\
          & \textit{IL} & 327.00   & Optimal  & 1496 \\
          & \textit{s-RLO} & 327.00 & Optimal  & 1496 \\
    \midrule
    \multirow{3}[2]{*}{TP-9a} & \textit{Relpscost} & 404.34 & Optimal  & 3672 \\
          & \textit{IL} & 318.77 & Optimal  & 1345 \\
          & \textit{s-RLO} & 281.94 & Optimal  & 266 \\
    \midrule
    \multirow{3}[2]{*}{TP-10a} & \textit{Relpscost} & 341.09 & Optimal  & 446 \\
          & \textit{IL} & 241.23 & Optimal  & 1623 \\
          & \textit{s-RLO} & 241.23 & Optimal  & 1623 \\
    \bottomrule
    \end{tabular}%
  \label{tab:tabel3}%
\end{table}%

Table \ref{tab:tabel3} presents a detailed comparison of the proposed \emph{s-RLO} approach against \emph{relpscost} and \emph{IL} on ten randomly selected test problems (TP-1a to TP-10a) with a 1440-h horizon. All methods successfully identify the optimal solution within the specified time limit, with \emph{s-RLO} consistently exhibiting the shortest solving time across all problems. On the majority of test problems, \emph{s-RLO} achieves significantly faster solving times than \emph{IL}. This improvement stems from \emph{RL} discovering a more efficient solving trajectory than \emph{IL}, that is, \emph{s-RLO} benefits from the superior performance of \emph{RL}. Notably, for TP-4a, TP-8a, and TP-10a, the performance of \emph{s-RLO} is equivalent to that of \emph{IL}. In these instances, \emph{RL} fails to identify a more efficient trajectory than \emph{IL}, and \emph{s-RLO} adopts the result of \emph{IL} with superior performance.

\begin{figure*}[!t]
\centering
\includegraphics[width=7in]{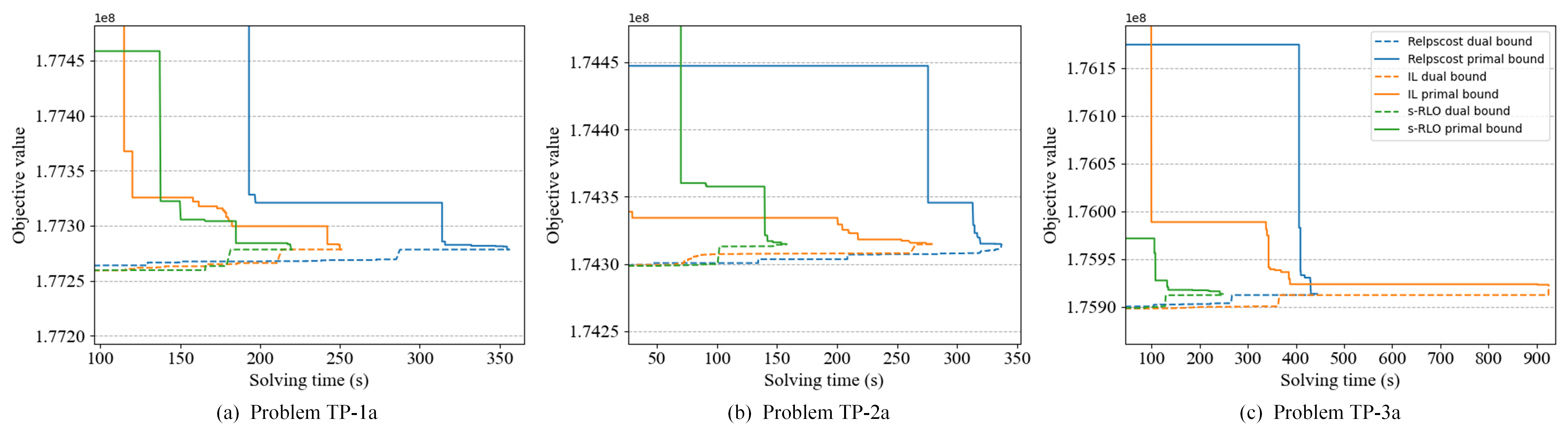}
\caption{Progress on dual and primal bounds for solving 1440-h PCM problems.}
\label{fig 7}
\end{figure*}

\begin{figure}[!t]
\centering
\includegraphics[width=3.6in]{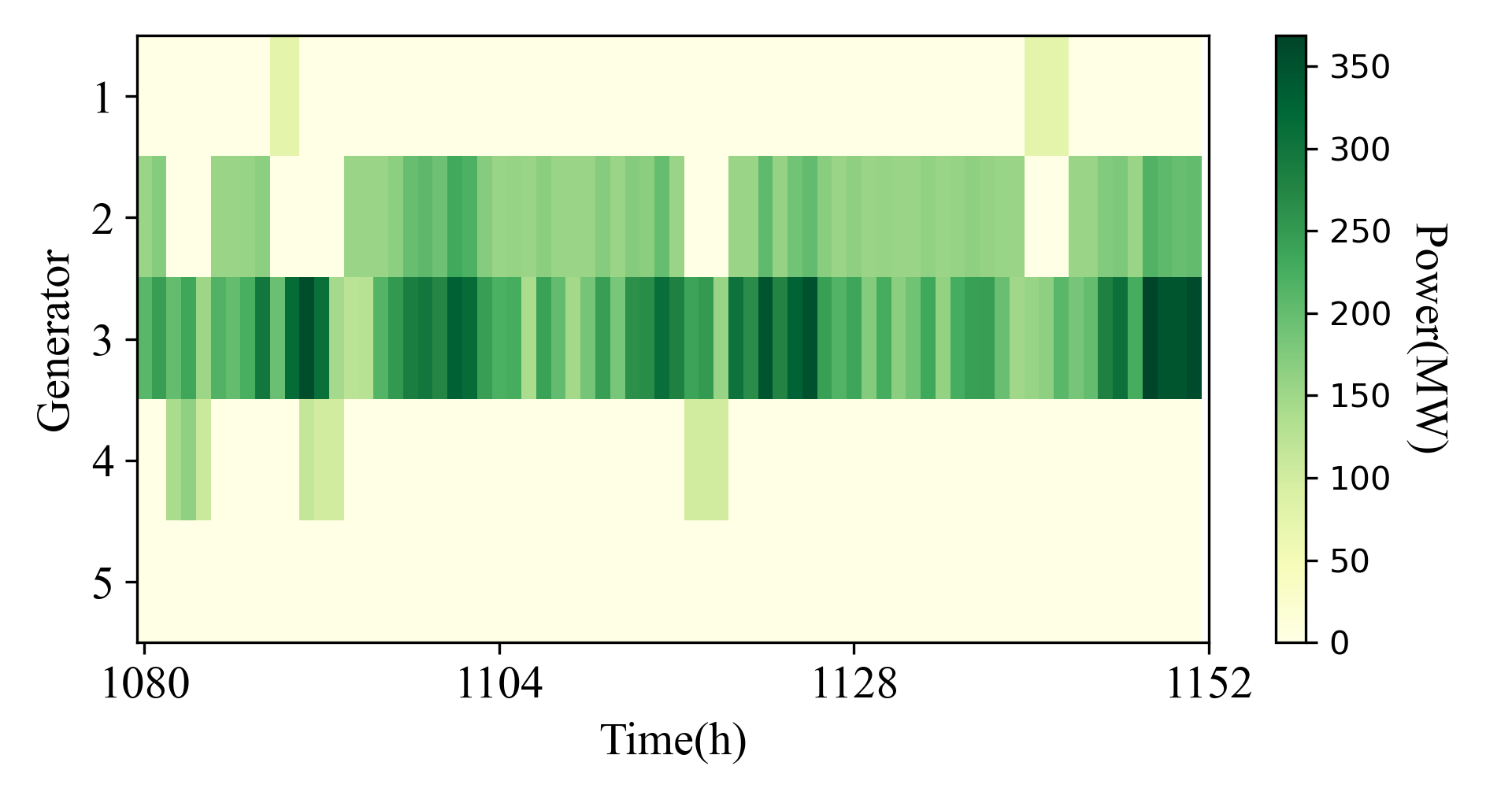}
\caption{Partial scheduling results for TP-2a.}
\label{fig 8}
\end{figure}

In various test problems, \emph{IL} exhibits a tendency to explore more nodes than \emph{relpscost} within a similar or shorter time. This suggests that using learned weight vectors to evaluate candidate variables is faster than relying on \emph{relpscost}. Although the variables chosen by \emph{IL} may not precisely align with those selected by \emph{relpscost}, and might not share the same level of quality, the speed advantage allows \emph{IL} to explore more nodes, leading to faster identification of optimal solutions. In some test problems, \emph{s-RLO} explores fewer nodes than both \emph{IL} and \emph{relpscost} in a shorter time. With the aid of \emph{RL}, which applies mini-batch rewards for updating the weight vector iteratively, \emph{s-RLO} discovers better solution trajectories than \emph{relpscost} and \emph{IL}. Consequently, the solving time is significantly reduced.

Taking problems TP-1a, TP-2a, and TP-3a as examples, Fig. \ref{fig 7} illustrates the progression of the dual and primal bounds using the different approaches. The proposed \emph{s-RLO}  consistently attains higher dual and lower primal bounds at an earlier point during the solution process than \emph{relpscost} and \emph{IL}. In particular, \emph{s-RLO} yields either a smaller gap in the same solving time or the same gap in a shorter solving time. The partial scheduling results for TP-2a, illustrated in Fig. \ref{fig 8}, verify that the constraints outlined in \eqref{eq2}--\eqref{eq14} are satisfied.

\subsection{Results Analysis on IEEE 118-Bus System }

The effectiveness of the proposed approach is further evaluated using an IEEE 118-bus system, featuring 54 generators, 186 branches, a wind farm situated at bus 26, and a solar farm located at bus 55. This system has an aggregate generation capacity of 9966.2 MW. Two distinct cases in the IEEE 118-bus system are established, with separate PCM horizons $T$ of 48 h and 168 h; the time interval is set to 1 h.

The load and renewable profiles are derived from actual data for a Chinese province, scaled down to 30\% for the experiments. A total of 1100 PCM problems are generated for each case by randomly multiplying the original profiles with sets of normal random noise. The sizes of the PCM problems for the 48- and 168-h horizons are listed in Table \ref{tab:table4}. The PCM problems are divided into 500 training problems for \emph{IL}, 500 training problems for \emph{RL}, and 100 test problems.

\begin{table}[!t]
\caption{Size of IEEE 118-bus system problems\label{tab:table4}}
\centering
\begin{tabular}{cccc}
\toprule
$T$ (h) & \# of Cont. var. & \# of Bin. var. & \# of Const.\\
\midrule
48 & 17376 & 2592 & 35328 \\
168 & 60816 & 9072 & 124728 \\
\bottomrule
\end{tabular}
\end{table}

\begin{figure}[t]
\centering
\includegraphics[width=3.2in]{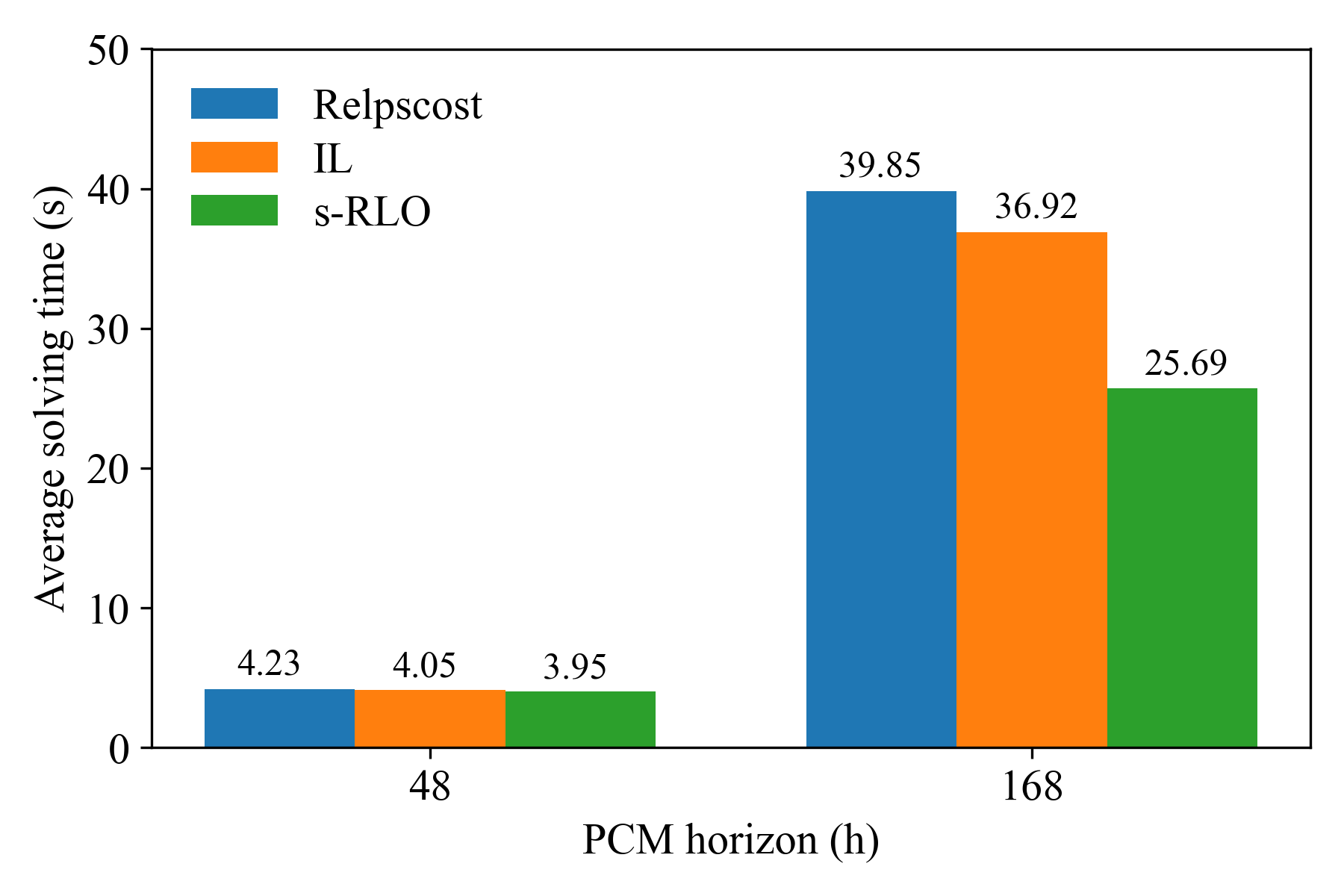}
\caption{Average solving times in IEEE 118-bus system.}
\label{fig 9}
\end{figure}

\begin{figure}[htb]
\centering
\includegraphics[width=3in]{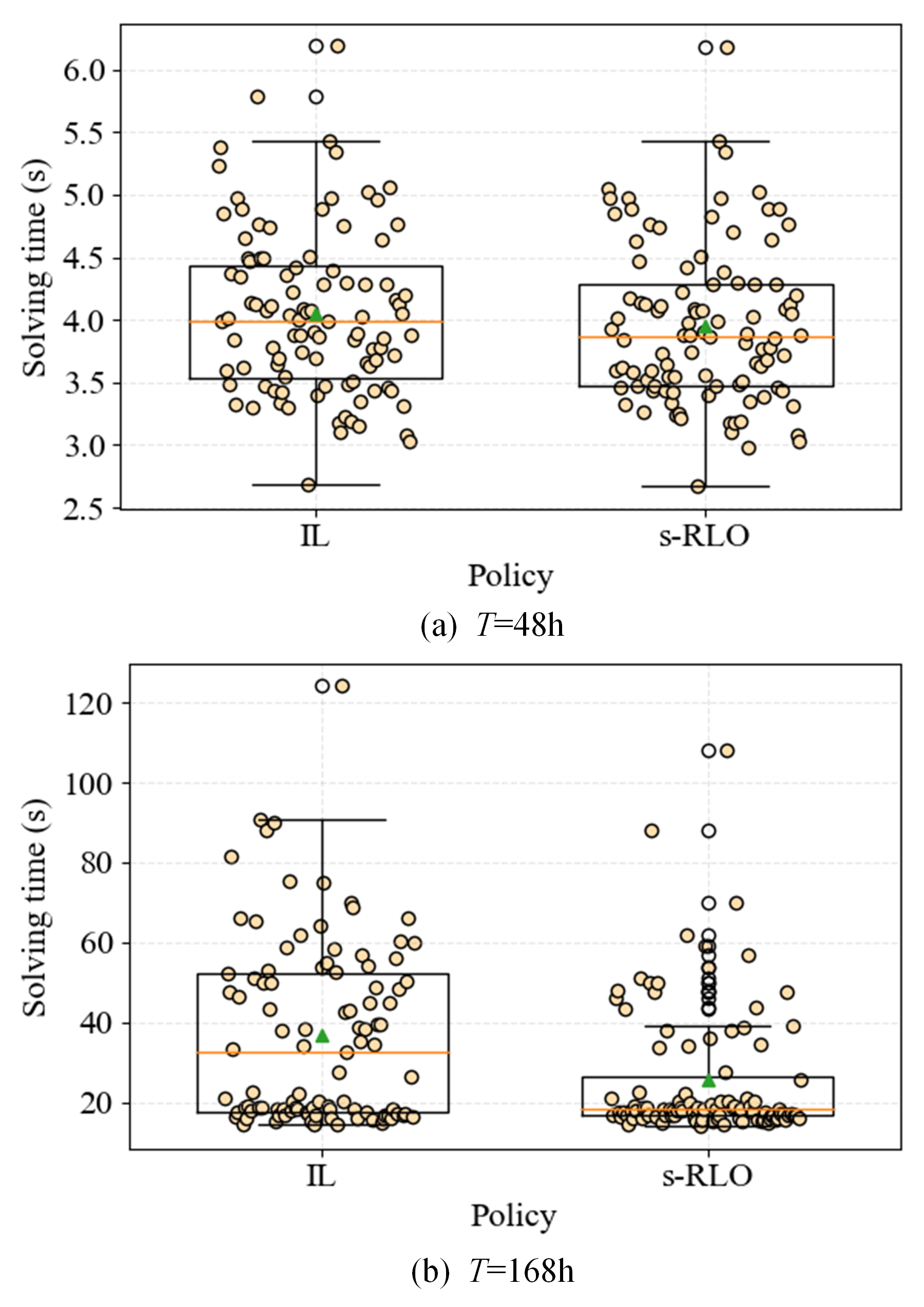}
\caption{Distribution of solving times in IEEE 118-bus system.}
\label{fig 10}
\end{figure}

Figure \ref{fig 9} illustrates the average solving times for the test problems in the IEEE 118-bus system. Applying \emph{IL} leads to a reduction of more than 5\% in the average solving time compared with \emph{relpscost}. The proposed \emph{s-RLO} approach further shortens the solving time over that of \emph{IL}. For the 168-h horizon problems, \emph{s-RLO} produces a reduction of more than 30\% in solving time compared with \emph{IL} and more than 35\% compared with \emph{relpscost}. In the current case, \emph{IL} does not significantly improve the solving speed of PCM problems. With the support of \emph{RL}, the solution time can be further reduced.

\begin{table}[!t]
  \centering
  \caption{Evaluation on randomly selected test problems}
    \begin{tabular}{ccccc}
    \toprule
   Problems & Approach & Solving time (s) & Status & $N_{EP}$ \\
    \midrule
    \multirow{3}[2]{*}{TP-1b} & \textit{Relpscost} & 30    & Optimal  & 43 \\
          & \textit{IL} & 14    & Optimal  & 24 \\
          & \textit{s-RLO} & 14    & Optimal  & 24 \\
    \midrule
    \multirow{3}[2]{*}{TP-2b} & \textit{Relpscost} & 25    & Optimal  & 41 \\
          & \textit{IL} & 27    & Optimal  & 214 \\
          & \textit{s-RLO} & 27    & Optimal  & 214 \\
    \midrule
    \multirow{3}[2]{*}{TP-3b} & \textit{Relpscost} & 357   & Optimal  & 4735 \\
          & \textit{IL} & 124   & Optimal  & 78 \\
          & \textit{s-RLO} & 108   & Optimal  & 14 \\
    \midrule
    \multirow{3}[2]{*}{TP-4b} & \textit{Relpscost} & 37    & Optimal  & 50 \\
          & \textit{IL} & 20    & Optimal  & 58 \\
          & \textit{s-RLO} & 20    & Optimal  & 58 \\
    \midrule
    \multirow{3}[2]{*}{TP-5b} & \textit{Relpscost} & 39    & Optimal  & 54 \\
          & \textit{IL} & 43    & Optimal  & 718 \\
          & \textit{s-RLO} & 38    & Optimal  & 456 \\
    \midrule
    \multicolumn{1}{c}{\multirow{3}[2]{*}{TP-6b}} & \textit{Relpscost} & 35    & Optimal  & 46 \\
          & \textit{IL} & 33    & Optimal  & 283 \\
          & \textit{s-RLO} & 19    & Optimal  & 46 \\
    \midrule
    \multicolumn{1}{c}{\multirow{3}[2]{*}{TP-7b}} & \textit{Relpscost} & 34    & Optimal  & 58 \\
          & \textit{IL} & 43    & Optimal  & 556 \\
          & \textit{s-RLO} & 20    & Optimal  & 71 \\
    \midrule
    \multicolumn{1}{c}{\multirow{3}[2]{*}{TP-8b}} & \textit{Relpscost} & 25    & Optimal  & 36 \\
          & \textit{IL} & 70    & Optimal  & 1301 \\
          & \textit{s-RLO} & 70   & Optimal  & 1301 \\
    \midrule
    \multicolumn{1}{c}{\multirow{3}[2]{*}{TP-9b}} & \textit{Relpscost} & 29    & Optimal  & 45 \\
          & \textit{IL} & 69    & Optimal  & 1384 \\
          & \textit{s-RLO} & 16    & Optimal  & 38 \\
    \midrule
    \multicolumn{1}{c}{\multirow{3}[2]{*}{TP-10b}} & \textit{Relpscost} & 29    & Optimal  & 51 \\
          & \textit{IL} & 18    & Optimal  & 76 \\
          & \textit{s-RLO} & 18    & Optimal  & 76 \\
    \bottomrule
    \end{tabular}%
  \label{tab:table5}%
\end{table}%

\begin{figure}[!t]
\centering
\includegraphics[width=3.2in]{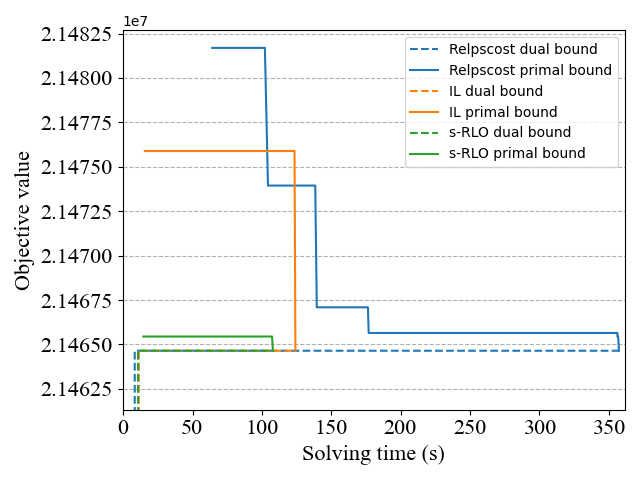}
\caption{Progress on dual and primal bounds for solving 168-h PCM problems.}
\label{fig 11}
\end{figure}

\begin{figure}[!t]
\centering
\includegraphics[width=3.9in]{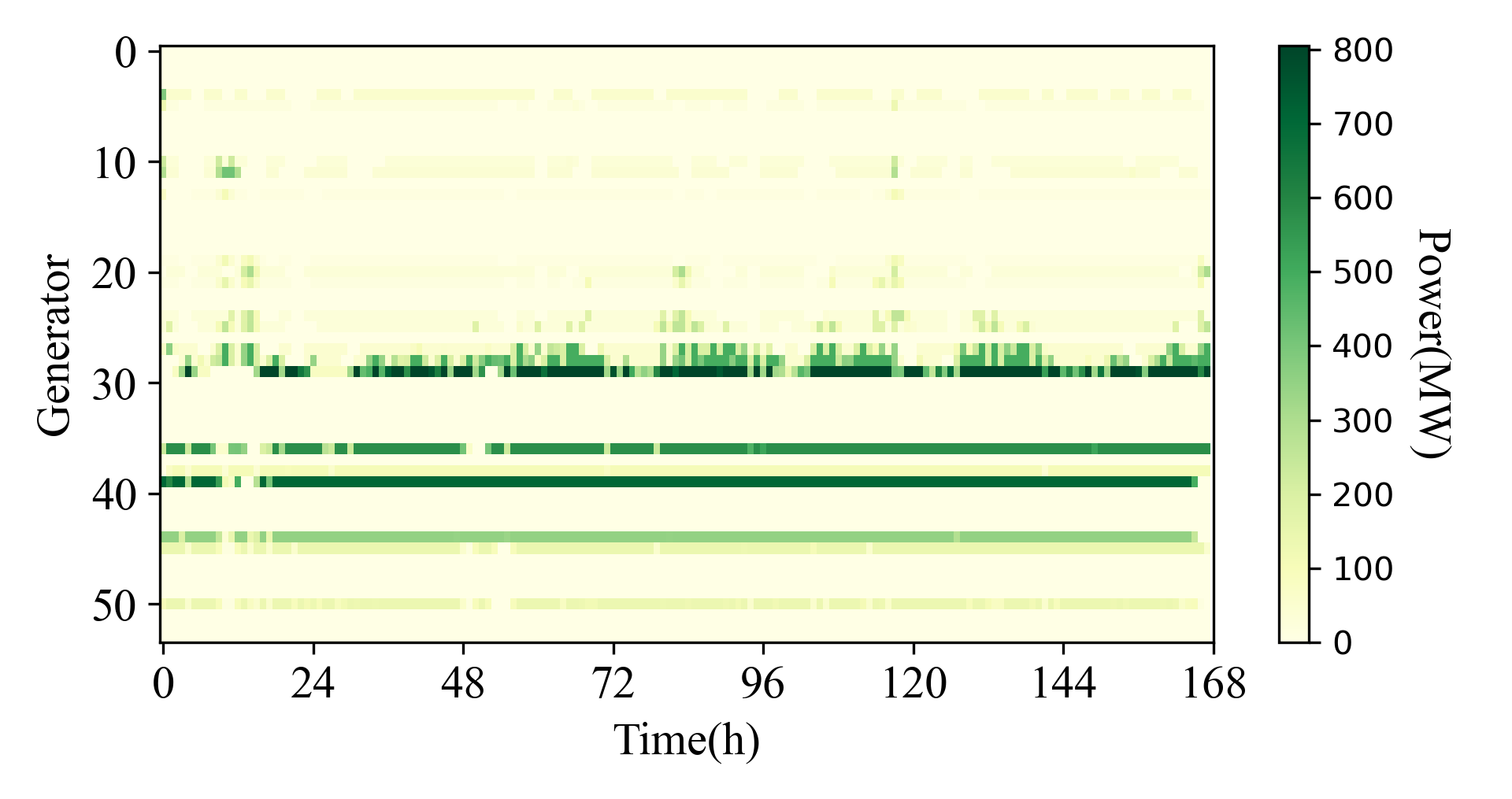}
\caption{Scheduling results for TP-3b in IEEE 118-bus system.}
\label{fig 12}
\end{figure}

In addition, the performance of \emph{ s-RLO} is more stable than that of \emph{IL} for these test problems. As shown in Figure \ref{fig 10}, applying \emph{IL} gives a more scattered distribution of solving times for the test PCMs than applying \emph{s-RLO}. For the 168-h horizon problems, the variance is 517.73 $\text{s}^{2}$ for \emph{IL} and 279.76 $\text{s}^{2}$ for \emph{s-RLO}. The variance of the solving time is reduced by more than 45\% by applying \emph{s-RLO}. In summary, \emph{s-RLO} has a more stable and superior acceleration effect than \emph{IL} on different scale problems, achieving 1.4$\times$ acceleration and reducing the variance of the solving time by approximately 2$\times$ compared with \emph{IL}.

Table \ref{tab:table5} presents a detailed comparison of the proposed \emph{s-RLO} approach against \emph{relpscost} and \emph{IL} on ten randomly selected test problems (TP-1b to TP-10b) with a 168-h horizon. All methods achieve optimal solutions within the time limit. \emph{s-RLO} demonstrates the shortest average solving time across all test problems. On problems TP-3b, TP-5b, TP-6b, TP-7b, and TP-9b, \emph{s-RLO} significantly outperforms \emph{IL}, benefiting from the ability of \emph{RL} to find better solution trajectories. In other problems, such as TP-1, \emph{s-RLO} exhibits equivalent performance to \emph{IL}, because \emph{RL} does not identify a superior trajectory and the \emph{IL} result is used by \emph{s-RLO}.

Typically, ML-selected variables differ from those chosen by \emph{relpscost}, and their quality may not match those of \emph{relpscost} because of the inherent limitation of ML accuracy. However, \emph{s-RLO} reduces the solving time while exploring more nodes, as variable selection with weight vectors is faster than that of \emph{relpscost}. Fast variable selection, even though ML, can lead to increased overall solving times with a large number of explored nodes. This highlights the importance of ensuring that ML-selected variables maintain good quality. On some test problems (such as TP-3b and TP-9b), \emph{s-RLO} exhibits better performance than \emph{relpscost} in terms of the number of nodes explored. Thus, it does not rule out the existence of accidental factors. Hence, \emph{s-RLO} may acquire some better knowledge than the expert \emph{relpscost} through \emph{RL} with the goal of minimizing the solving time.

Taking TP-3b as an example, Figure \ref{fig 11} illustrates the progress of the dual and primal bounds using the different approaches. The proposed \emph{s-RLO} achieves a lower primal bound at an earlier time during the solving progress than both \emph{relpscost} and \emph{IL}. Specifically, \emph{s-RLO} either provides a smaller gap in the same solving time or the same gap in a shorter solving time. Figure \ref{fig 12} presents the scheduling results for TP-3b, demonstrating that the constraints outlined in \eqref{eq2}--\eqref{eq14} are satisfied.

\section{Conclusion}
\label{section 5}
This paper introduces a novel approach to solving PCM problems, which are traditionally plagued by NP-hard complexity and a multitude of binary variables. Modern power systems demand rapid solutions to PCM problems to allow swift evaluation of diverse planning options. Although existing methods often prioritize speed at the expense of optimality, our work employs an ML-enhanced B\&B algorithm to achieve both speed and optimality. We described the novel \emph{s-RLO} approach, which seamlessly integrates \emph{IL} for rapid policy network formation with \emph{RL} for subsequent policy network refinement. Empirical results demonstrated that our approach outperforms existing techniques such as \emph{relpscost} and \emph{IL}. Specifically, the solution speed is up to 2$\times$ faster while maintaining solution optimality, and the time variance of the solving process is reduced by up to 50\%. These advances make our approach an important tool for reliable and efficient production cost planning in ever-evolving power systems.

Our work can be extended to other complicated PCM and MILP problems in power systems. Further explorations will consider applications to more complicated PCM problems focusing on uncertainty and resilience, and attempt to combine \emph{s-RLO} with other acceleration methods such as the time-domain partition-based approach.

\section*{Acknowledgments}
This work is supported by the Natural Science Foundation of China [grant number 52277086 and 52307085]; and Natural Science Foundation of Jiangsu Province [grant number BK20230027].


\newpage

 






\begin{thebibliography}{1}
\bibliographystyle{IEEEtran}

\bibitem{ref1}
N. Zhang, H. Jiang, E. Du, Z. Zhuo, P. Wang, Z. Wang and Y. Zhang, “An Efficient Power System Planning Model Considering Year-round Hourly Operation Simulation,” {\it {IEEE Trans. Power Syst.}}, vol. 37, no. 6, pp. 4925-4935, 2022.

\bibitem{ref2}
X. Li, Q. Zhai, J. Zhou and X. Guan, “A Variable Reduction Method for Large-Scale Unit Commitment,” {\it {IEEE Trans. Power Syst.}}, vol. 35, no. 1, pp. 261–272, 2020.

\bibitem{ref reduction1}
M. Zhang, Z. Yang, W. Lin, J. Yu, W. Dai and E. Du, “Enhancing economics of power systems through fast unit commitment with high time resolution,” {\it {Appl. Energy}}, vol. 281, pp. 116051, 2021.

\bibitem{ref reduction2}
M.Eslami, H. A. Moghadam, H. Zayandehroodi and N. Ghadimi, “A new formulation to reduce the number of variables and constraints to expedite SCUC in bulky power systems,” {\it {Proceedings of the national academy of sciences, india section a: physical sciences}}, vol. 89, pp. 311-321, 2019.

\bibitem{ref reduction3}
Y. Yin, C. He, T. Liu and L. Wu, “Risk-Averse Stochastic Midterm Scheduling of Thermal-Hydro-Wind System: A Network-Constrained Clustered Unit Commitment Approach,” {\it {IEEE. Sustain. Energy}}, vol. 13, no. 3, pp. 1293-1304, 2022.

\bibitem{ref reduction4}
J. Meus, K. Poncelet and E. Delarue, “Applicability of a Clustered Unit Commitment Model in Power System Modeling,” {\it {IEEE Trans. Power Syst.}}, vol. 33, no. 2, pp. 2195-2204, 2018.

\bibitem{ref reduction5}
C. Feng, C. Shao, X. Wang, “CSP clustering in unit commitment for power system production cost modeling,” {\it {Renew. Energ.}}, vol. 168, pp. 1217-1228, 2021.

\bibitem{ref reduction6}
N. Zhang, H. Jiang, E. Du, Z. Zhuo, P. Wang, Z. Wang, Y. Zhang, “An Efficient Power System Planning Model Considering Year-Round Hourly Operation Simulation,” {\it {IEEE Trans. Power Syst.}}, vol. 37, no. 6, pp. 4925-4935, 2022.

\bibitem{ref relxation1}
B. Hua, R. Baldick and J. Wang, “Representing Operational Flexibility in Generation Expansion Planning Through Convex Relaxation of Unit Commitment,” {\it {IEEE Trans. Power Syst.}}, vol. 33, no. 2, pp. 2272-2281, 2018.

\bibitem{ref relxation2}
X. Sun, P. B. Luh, M. A. Bragin, Y. Chen, J. Wan and F. Wang, “A Novel Decomposition and Coordination Approach for Large Day-Ahead Unit Commitment With Combined Cycle Units,” {\it {IEEE Trans. Power Syst.}}, vol. 33, no. 5, pp. 5297-5308, 2018.

\bibitem{ref relxation3}
M. Qu, T. Ding, C. Mu, X. Zhang, K. Pan and M. Shahidehpour, “Linearization Method for Large-Scale Hydro-Thermal Security-Constrained Unit Commitment,” {\it {IEEE Trans. Autom. Sci. Eng.}}, Early Access.

\bibitem{ref partition1}
C. Barrows, M. Hummon, W. Jones, and E. Hale, “Time domain partitioning of electricity production cost simulations,”  United States: N. p., 2014.

\bibitem{ref partition2}
T. Xu and N. Zhang, “Coordinated Operation of Concentrated Solar Power and Wind Resources for the Provision of Energy and Reserve Services,” {\it {IEEE Trans. Power Syst.}}, vol. 32, no. 2, pp. 1260–1271, 2017.

\bibitem{ref partition3}
S. Malekshah and J. Ansari, “A novel decentralized method based on the system engineering concept for reliability‐security constraint unit commitment in restructured power environment,” {\it {Int. J. Energy Res.}}, vol. 45, no. 1,  pp. 703-726, 2021.


\bibitem{ref ML1-1}
S. Gao, C. Xiang, M. Yu, K. T. Tan and T. H. Lee, “Online Optimal Power Scheduling of a Microgrid via Imitation Learning,” {\it {IEEE Trans Smart Grid}}, vol. 13, no. 2, pp. 861-876, 2022.

\bibitem{ref ML2-1}
T. Wu, Y. J. Angela Zhang, and S. Wang, “Deep learning to optimize: security-constrained unit commitment with uncertain wind power generation and BESSs,” {\it {IEEE Trans. Sustain. Energy}}, vol. 13, no. 1, pp.231-240, 2022.
\bibitem{ref ML2-2}
L. Sang, Y. Xu and H. Sun, “Ensemble Provably Robust Learn-to-Optimize Approach for Security-Constrained Unit Commitment,” {\it {IEEE Trans. Power Syst.}}, Early Access.

\bibitem{ref ML3-1}
F. Mohammadi, M. Sahraei-Ardakani, D. N. Trakas and N. D. Hatziargyriou, “Machine Learning Assisted Stochastic Unit Commitment During Hurricanes With Predictable Line Outages,” {\it {IEEE Trans. Power Syst.}}, vol. 36, no. 6, pp. 5131-5142, 2021.

\bibitem{ref ML3-2}
A. V. Ramesh and X. Li, “Machine Learning Assisted Model Reduction for Security-Constrained Unit Commitment,” 2022 North American Power Symposium (NAPS), Salt Lake City, UT, USA, pp. 1-6, 2022.

\bibitem{ref3}
K. Kim, A. Botterud and F. Qiu, “Temporal Decomposition for Improved Unit Commitment in Power System Production Cost Modeling,” {\it {IEEE Trans. Power Syst.}}, vol. 33, no. 5, pp. 5276-5287, 2018.

\bibitem{ref3.5}
G. Gamrath, T. Fischer, T. Gally, A. M. Gleixner, G. Hendel, T. Koch, S. J. Maher, M. Miltenberger, B. Muller, M. E. Pfetsch, ¨C. Puchert, D. Rehfeldt, S. Schenker, R. Schwarz, F. Serrano, Y. Shinano, S. Vigerske, D. Weninger, M. Winkler, J. T. Witt, and J. Witzig, “The SCIP Optimization Suite 3.2,” Takustr.7, 14195 Berlin, Tech. Rep., 2016.

\bibitem{ref BB-1}
L Huang, X Chen, W Huo, J Wang, F Zhang, B Bai and L Shi, “Branch and bound in mixed integer linear programming problems: A survey of techniques and trends,” arXiv preprint arXiv:2111.06257, 2021.

\bibitem{ref BB-2}
V. Nair, S. Bartunov, F. Gimeno, et al, “Solving mixed integer programs using neural networks,” arXiv preprint arXiv:2012.13349, 2020.

\bibitem{ref BB-3}
A. Lodi and G. Zarpellon, “On learning and branching: a survey,” {\it {Top}}, vol. 25, pp. 207-236, 2017.

\bibitem{ref BB-4}
M. Huang, L. Huang, Y. Zhong, et al, “MILP Acceleration: A Survey from Perspectives of Simplex Initialization and Learning-Based Branch and Bound,” {\it {J. Oper. Res. Soc. China}}, pp. 1-55, 2023.


\bibitem{ref4}
Q. Hu, Z. Guo and F. Li, “Imitation Learning Based Fast Power System Production Cost Minimization Simulation,” {\it {IEEE Trans. Power Syst.}}, vol. 38, no. 3, pp. 2951-2954, 2023.


\bibitem{ref5}
H. He, H. Daume III and JM Eisner, “Learning to search in branch and bound algorithms,” {\it {Proc. Adv. Neural Inf. Process. Syst.}}, vol. 27, 2014.

\bibitem{ref6}
R. Howard, {\it {Dynamic Programming and Markov Processes}}. Cambridge, MA: MIT Press, 1960.

\bibitem{ref7}
G. Zarpellon, J. Jo, A. Lodi, and Y. Bengio, “Parameterizing Branchand-Bound Search Trees to Learn Branching Policies,” AAAI, vol. 35, no.5, pp. 3931–3939, 2021.

\end{thebibliography}
\end{document}